\documentclass[%
reprint,
superscriptaddress,
showpacs,
 amsmath,amssymb,
 aps,
 prl,
]{revtex4-1}

\usepackage{graphicx}
\usepackage{dcolumn}
\usepackage{bm}
\usepackage{natbib}
\usepackage{multirow}

\def\csnospin{\sigma_{\chi N}^{\rm{SI}}}

\begin{document}

\preprint{APS/123-QED}

\title{Limits on Light Weakly Interacting Massive Particles from the First 102.8 kg ${\times}$ day Data of the CDEX-10 Experiment}

\affiliation{Key Laboratory of Particle and Radiation Imaging (Ministry of Education) and Department of Engineering Physics, Tsinghua University, Beijing 100084}
\affiliation{College of Nuclear Science and Technology, Beijing Normal University, Beijing 100875}
\affiliation{Institute of Physics, Academia Sinica, Taipei 11529}
\affiliation{Department of Physics, Dokuz Eyl\"{u}l University, \.{I}zmir 35160}
\affiliation{Department of Physics, Tsinghua University, Beijing 100084}
\affiliation{NUCTECH Company, Beijing 100084}
\affiliation{YaLong River Hydropower Development Company, Chengdu 610051}
\affiliation{College of Physical Science and Technology, Sichuan University, Chengdu 610064}
\affiliation{Department of Nuclear Physics, China Institute of Atomic Energy, Beijing 102413}
\affiliation{School of Physics, Nankai University, Tianjin 300071}
\affiliation{Department of Physics, Banaras Hindu University, Varanasi 221005}
\affiliation{Department of Physics, Beijing Normal University, Beijing 100875}
\author{H. Jiang}
\affiliation{Key Laboratory of Particle and Radiation Imaging (Ministry of Education) and Department of Engineering Physics, Tsinghua University, Beijing 100084}
\author{L.~P.~Jia}
\affiliation{Key Laboratory of Particle and Radiation Imaging (Ministry of Education) and Department of Engineering Physics, Tsinghua University, Beijing 100084}
\author{Q. Yue}\altaffiliation [Corresponding author: ]{yueq@mail.tsinghua.edu.cn}
\affiliation{Key Laboratory of Particle and Radiation Imaging (Ministry of Education) and Department of Engineering Physics, Tsinghua University, Beijing 100084}
\author{K.~J.~Kang}
\affiliation{Key Laboratory of Particle and Radiation Imaging (Ministry of Education) and Department of Engineering Physics, Tsinghua University, Beijing 100084}
\author{J.~P.~Cheng}
\affiliation{Key Laboratory of Particle and Radiation Imaging (Ministry of Education) and Department of Engineering Physics, Tsinghua University, Beijing 100084}
\affiliation{College of Nuclear Science and Technology, Beijing Normal University, Beijing 100875}
\author{Y.~J.~Li}
\affiliation{Key Laboratory of Particle and Radiation Imaging (Ministry of Education) and Department of Engineering Physics, Tsinghua University, Beijing 100084}
\author{H.~T.~Wong}
\altaffiliation{Participating as a member of TEXONO Collaboration}
\affiliation{Institute of Physics, Academia Sinica, Taipei 11529}
\author{M. Agartioglu}
\altaffiliation{Participating as a member of TEXONO Collaboration}
\affiliation{Institute of Physics, Academia Sinica, Taipei 11529}
\affiliation{Department of Physics, Dokuz Eyl\"{u}l University, \.{I}zmir 35160}
\author{H.~P.~An}
\affiliation{Key Laboratory of Particle and Radiation Imaging (Ministry of Education) and Department of Engineering Physics, Tsinghua University, Beijing 100084}
\affiliation{Department of Physics, Tsinghua University, Beijing 100084}
\author{J.~P.~Chang}
\affiliation{NUCTECH Company, Beijing 100084}
\author{J.~H.~Chen}
\altaffiliation{Participating as a member of TEXONO Collaboration}
\affiliation{Institute of Physics, Academia Sinica, Taipei 11529}
\author{Y.~H.~Chen}
\affiliation{YaLong River Hydropower Development Company, Chengdu 610051}
\author{Z.~Deng}
\affiliation{Key Laboratory of Particle and Radiation Imaging (Ministry of Education) and Department of Engineering Physics, Tsinghua University, Beijing 100084}
\author{Q.~Du}
\affiliation{College of Physical Science and Technology, Sichuan University, Chengdu 610064}
\author{H.~Gong}
\affiliation{Key Laboratory of Particle and Radiation Imaging (Ministry of Education) and Department of Engineering Physics, Tsinghua University, Beijing 100084}
\author{L. He}
\affiliation{NUCTECH Company, Beijing 100084}
\author{J.~W.~Hu}
\affiliation{Key Laboratory of Particle and Radiation Imaging (Ministry of Education) and Department of Engineering Physics, Tsinghua University, Beijing 100084}
\author{Q.~D.~Hu}
\affiliation{Key Laboratory of Particle and Radiation Imaging (Ministry of Education) and Department of Engineering Physics, Tsinghua University, Beijing 100084}
\author{H.~X.~Huang}
\affiliation{Department of Nuclear Physics, China Institute of Atomic Energy, Beijing 102413}
\author{H.~B.~Li}
\altaffiliation{Participating as a member of TEXONO Collaboration}
\affiliation{Institute of Physics, Academia Sinica, Taipei 11529}
\author{H. Li}
\affiliation{NUCTECH Company, Beijing 100084}
\author{J.~M.~Li}
\affiliation{Key Laboratory of Particle and Radiation Imaging (Ministry of Education) and Department of Engineering Physics, Tsinghua University, Beijing 100084}
\author{J.~Li}
\affiliation{Key Laboratory of Particle and Radiation Imaging (Ministry of Education) and Department of Engineering Physics, Tsinghua University, Beijing 100084}
\author{X.~Li}
\affiliation{Department of Nuclear Physics, China Institute of Atomic Energy, Beijing 102413}
\author{X.~Q.~Li}
\affiliation{School of Physics, Nankai University, Tianjin 300071}
\author{Y.~L.~Li}
\affiliation{Key Laboratory of Particle and Radiation Imaging (Ministry of Education) and Department of Engineering Physics, Tsinghua University, Beijing 100084}

\author {B. Liao}
\affiliation{College of Nuclear Science and Technology, Beijing Normal University, Beijing 100875}

\author{F.~K.~Lin}
\altaffiliation{Participating as a member of TEXONO Collaboration}
\affiliation{Institute of Physics, Academia Sinica, Taipei 11529}
\author{S.~T.~Lin}
\affiliation{College of Physical Science and Technology, Sichuan University, Chengdu 610064}
\author{S.~K.~Liu}
\affiliation{College of Physical Science and Technology, Sichuan University, Chengdu 610064}

\author {Y.~D.~Liu}
\affiliation{College of Nuclear Science and Technology, Beijing Normal University, Beijing 100875}
\author {Y.~Y.~Liu}
\affiliation{College of Nuclear Science and Technology, Beijing Normal University, Beijing 100875}

\author{Z.~Z.~Liu}
\affiliation{Key Laboratory of Particle and Radiation Imaging (Ministry of Education) and Department of Engineering Physics, Tsinghua University, Beijing 100084}
\author{H.~Ma}\altaffiliation [Corresponding author: ]{mahao@mail.tsinghua.edu.cn}
\affiliation{Key Laboratory of Particle and Radiation Imaging (Ministry of Education) and Department of Engineering Physics, Tsinghua University, Beijing 100084}
\author{J.~L.~Ma}
\affiliation{Key Laboratory of Particle and Radiation Imaging (Ministry of Education) and Department of Engineering Physics, Tsinghua University, Beijing 100084}
\affiliation{Department of Physics, Tsinghua University, Beijing 100084}
\author{H.~Pan}
\affiliation{NUCTECH Company, Beijing 100084}
\author{J.~Ren}
\affiliation{Department of Nuclear Physics, China Institute of Atomic Energy, Beijing 102413}
\author{X.~C.~Ruan}
\affiliation{Department of Nuclear Physics, China Institute of Atomic Energy, Beijing 102413}
\author{B. Sevda}
\altaffiliation{Participating as a member of TEXONO Collaboration}
\affiliation{Institute of Physics, Academia Sinica, Taipei 11529}
\affiliation{Department of Physics, Dokuz Eyl\"{u}l University, \.{I}zmir 35160}
\author{V.~Sharma}
\altaffiliation{Participating as a member of TEXONO Collaboration}
\affiliation{Institute of Physics, Academia Sinica, Taipei 11529}
\affiliation{Department of Physics, Banaras Hindu University, Varanasi 221005}
\author{M.~B.~Shen}
\affiliation{YaLong River Hydropower Development Company, Chengdu 610051}
\author{L.~Singh}
\altaffiliation{Participating as a member of TEXONO Collaboration}
\affiliation{Institute of Physics, Academia Sinica, Taipei 11529}
\affiliation{Department of Physics, Banaras Hindu University, Varanasi 221005}
\author{M.~K.~Singh}
\altaffiliation{Participating as a member of TEXONO Collaboration}
\affiliation{Institute of Physics, Academia Sinica, Taipei 11529}
\affiliation{Department of Physics, Banaras Hindu University, Varanasi 221005}

\author {T.~X.~Sun}
\affiliation{College of Nuclear Science and Technology, Beijing Normal University, Beijing 100875}

\author{C.~J.~Tang}
\affiliation{College of Physical Science and Technology, Sichuan University, Chengdu 610064}
\author{W.~Y.~Tang}
\affiliation{Key Laboratory of Particle and Radiation Imaging (Ministry of Education) and Department of Engineering Physics, Tsinghua University, Beijing 100084}
\author{Y.~Tian}
\affiliation{Key Laboratory of Particle and Radiation Imaging (Ministry of Education) and Department of Engineering Physics, Tsinghua University, Beijing 100084}

\author {G.~F.~Wang}
\affiliation{College of Nuclear Science and Technology, Beijing Normal University, Beijing 100875}

\author{J.~M.~Wang}
\affiliation{YaLong River Hydropower Development Company, Chengdu 610051}
\author{L.~Wang}
\affiliation{Department of Physics, Beijing Normal University, Beijing 100875}
\author{Q.~Wang}
\affiliation{Key Laboratory of Particle and Radiation Imaging (Ministry of Education) and Department of Engineering Physics, Tsinghua University, Beijing 100084}
\affiliation{Department of Physics, Tsinghua University, Beijing 100084}
\author{Y.~Wang}
\affiliation{Key Laboratory of Particle and Radiation Imaging (Ministry of Education) and Department of Engineering Physics, Tsinghua University, Beijing 100084}
\affiliation{Department of Physics, Tsinghua University, Beijing 100084}
\author{S.~Y.~Wu}
\affiliation{YaLong River Hydropower Development Company, Chengdu 610051}
\author{Y.~C.~Wu}
\affiliation{Key Laboratory of Particle and Radiation Imaging (Ministry of Education) and Department of Engineering Physics, Tsinghua University, Beijing 100084}
\author{H.~Y.~Xing}
\affiliation{College of Physical Science and Technology, Sichuan University, Chengdu 610064}
\author{Y.~Xu}
\affiliation{School of Physics, Nankai University, Tianjin 300071}
\author{T.~Xue}
\affiliation{Key Laboratory of Particle and Radiation Imaging (Ministry of Education) and Department of Engineering Physics, Tsinghua University, Beijing 100084}
\author{L.~T.~Yang}\altaffiliation [Corresponding author: ]{yanglt@mail.tsinghua.edu.cn}
\affiliation{Key Laboratory of Particle and Radiation Imaging (Ministry of Education) and Department of Engineering Physics, Tsinghua University, Beijing 100084}
\affiliation{Department of Physics, Tsinghua University, Beijing 100084}
\author{S.~W.~Yang}
\altaffiliation{Participating as a member of TEXONO Collaboration}
\affiliation{Institute of Physics, Academia Sinica, Taipei 11529}
\author{N.~Yi}
\affiliation{Key Laboratory of Particle and Radiation Imaging (Ministry of Education) and Department of Engineering Physics, Tsinghua University, Beijing 100084}
\author{C.~X.~Yu}
\affiliation{School of Physics, Nankai University, Tianjin 300071}
\author{H.~J.~Yu}
\affiliation{NUCTECH Company, Beijing 100084}
\author{J.~F.~Yue}
\affiliation{YaLong River Hydropower Development Company, Chengdu 610051}
\author{X.~H.~Zeng}
\affiliation{YaLong River Hydropower Development Company, Chengdu 610051}
\author{M.~Zeng}
\affiliation{Key Laboratory of Particle and Radiation Imaging (Ministry of Education) and Department of Engineering Physics, Tsinghua University, Beijing 100084}
\author{Z.~Zeng}
\affiliation{Key Laboratory of Particle and Radiation Imaging (Ministry of Education) and Department of Engineering Physics, Tsinghua University, Beijing 100084}

\author {F.~S.~Zhang}
\affiliation{College of Nuclear Science and Technology, Beijing Normal University, Beijing 100875}

\author{Y.~H.~Zhang}
\affiliation{YaLong River Hydropower Development Company, Chengdu 610051}
\author{M.~G.~Zhao}
\affiliation{School of Physics, Nankai University, Tianjin 300071}
\author{J.~F.~Zhou}
\affiliation{YaLong River Hydropower Development Company, Chengdu 610051}
\author{Z.~Y.~Zhou}
\affiliation{Department of Nuclear Physics, China Institute of Atomic Energy, Beijing 102413}
\author{J.~J.~Zhu}
\affiliation{College of Physical Science and Technology, Sichuan University, Chengdu 610064}
\author{Z.~H.~Zhu}
\affiliation{YaLong River Hydropower Development Company, Chengdu 610051}

\collaboration{CDEX Collaboration}
\noaffiliation

\date{\today}

\begin{abstract}
We report the first results of a light weakly interacting massive particles (WIMPs) search from the CDEX-10 experiment with a 10 kg germanium detector array immersed in liquid nitrogen at the China Jinping Underground Laboratory with a physics data size of 102.8 kg day. At an analysis threshold of 160 eVee, improved limits of 8 $\times 10^{-42}$ and 3 $\times 10^{-36}$ cm$^{2}$  at a 90\% confidence level on spin-independent and spin-dependent WIMP-nucleon cross sections, respectively, at a WIMP mass ($m_{\chi}$) of 5 GeV/${c}^2$ are achieved. The lower reach of $m_{\chi}$ is extended to 2 GeV/${c}^2$.
\begin{description}
\item[PACS numbers]{95.35.+d,
29.40.-n,
98.70.Vc}
\end{description}
\end{abstract}

\maketitle


Weakly interacting massive particles (WIMPs, denoted as $\chi$) have been extensively searched via elastic scattering with normal matter in underground direct detection experiments~\cite{PDG2017,tech} under ultralow background conditions. Liquid noble gas detectors are leading the sensitivities at WIMP mass ($m_\chi$)  above 10 GeV/${c}^2$ \cite{lux,pandax,xenon}, while solid state detectors are generally used for ${m}_\chi~<~10~$ GeV/$c^2$ \cite{cogent,cdex0,cdex12014,cdex12016,cdex12018,supercdms,cdmslite,CRESST-II}.

With excellent energy resolution and low energy threshold, \emph{p}-type point contact germanium (\emph{p}PCGe) detectors have been used and further developed for light WIMP searches by CDEX \cite{cdex0,cdex12014,cdex12016,cdex12018}. Located in the China Jinping Underground Laboratory (CJPL)~\cite{cjpl}, the first generation \mbox{CDEX-1A (1B)} experiments used 1-kg-scale single-element $p$PCGe cooled by a cold finger since 2010 \cite{cdex12014,cdex12016,cdex12018}. With an energy threshold of 160 eVee (``eVee" represents electron equivalent energy derived from a charge calibration) and an exposure of 737.1 kg day, CDEX-1B provided improved limits on WIMP-nucleon spin-independent (SI) and spin-dependent (SD) scattering down to $m_{\chi}$ of 2 GeV/$c^2$~\cite{cdex12018}.

Toward a future ton-scale DM experiment, the second generation CDEX experiment with a total detector mass of about 10 kg, called CDEX-10, has used three triple-element $p$PCGe strings (C10A, B, C) directly immersed in liquid nitrogen (LN$_{2}$). Compared with cold finger cooling and high-$Z$ material shielding systems, low-$Z$ material shielding, such as with LN$_{2}$ or liquid argon, provides better control of radiation background. The concept of integrated shielding and cooling, first proposed in the GENIUS project~\cite{GENIUS}, is realized in the GERDA experiment with the lowest background among neutrinoless double beta decay (0$\nu\beta\beta$) experiments~\cite{gerda} and will be expanded into the next generation LEGEND 0$\nu\beta\beta$ program~\cite{LEGEND}. 
CDEX-10 focuses on the arraying technologies and background understanding of the prototype $p$PCGe detectors developed based on the CDEX-1 technique. The new CDEX-10 array detectors and dedicated data acquisition (DAQ) system started testing and data taking inside a LN$_{2}$ tank in 2016 at CJPL. C10A was returned to the CANBERRA factory in France for upgrades. Of the remaining six detectors, two had faulty cabling, and two others had a high level of noise. In this Letter, we report the results from a first physics data set of one of the two operational detectors C10B-Ge1, which had the lower threshold. 

\begin{figure}[h]
\includegraphics[width=0.9\linewidth]{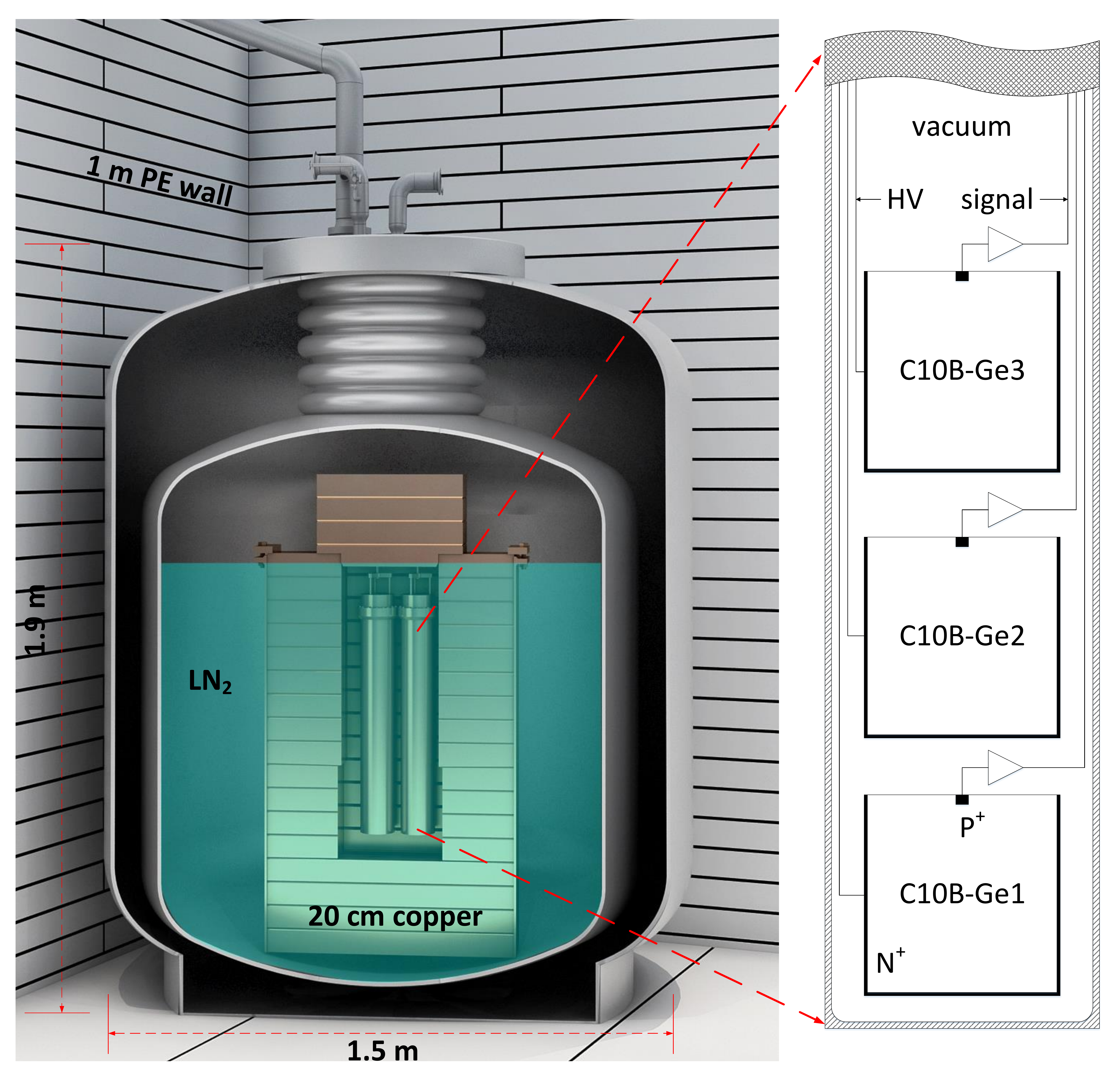}
\caption{
Configuration of CDEX-10 experimental setup (left) and C10B detector layout inside the string (right). C10B and C10C are running inside the LN$_{2}$ tank which has an outer diameter of 1.5 m and a height of 1.9 m. Each detector string consists of three PCGe detectors tagged as Ge1 to Ge3 from bottom to top. The size of each germanium crystal is approximately {$\Phi$}62 mm $\times$ H62 mm.
}
\label{fig::detector_config}
\end{figure}

The stainless steel LN$_{2}$ tank was located in the polyethylene room with 1 m thick walls at CJPL-I for cooling of the CDEX-10 detectors, which are surrounded by 20 cm thick high-purity oxygen-free copper immersed in LN$_{2}$ to shield the ambient radioactivities. The shielding configuration of CDEX-10 and the structure of a detector string are shown in Fig.~\ref{fig::detector_config}. 

The DAQ system received signals from the $p^{+}$ point contact electrode of C10B-Ge1 which were fed into a pulsed reset preamplifier. Five identical output signals of the preamplifier were further processed and digitized. Two of them were distributed into 6 $\mu$s ($S_{p6}$) and 12 $\mu$s ($S_{p12}$) shaping amplifiers for a 0-12 keVee energy range. These two channels were used for energy calibration and signal and noise discrimination. The third channel was loaded to a timing amplifier ($T_{p}$) to measure the rise time of signals within a 0-12 keVee energy range which can be used for bulk or surface events discrimination. The remaining two were loaded to a shaping amplifier and a timing amplifier with low gains aiming at a high energy range for background understanding. To estimate the dead time of the DAQ system and cut efficiencies uncorrelated with energies, random trigger (RT) events were recorded once every 20 seconds. The output signals of the above amplifiers were digitized by the 14-bit 100-MHz flash analog-to-digital converters. Data taking with C10B-Ge1 was performed from February 26, 2017 to November 7, 2017. The DAQ dead time fraction was measured by RT events to be 4.8\%, giving a live time of 112.3 days.

The data analysis follows the procedures described in our earlier work \cite{cdex12014,cdex12016,cdex12018}, starting from the parameters extracted from the digitized pulses. The optimal integrated area of the pulse from $S_{p12}$ is selected to define the energy for its excellent energy linearity at the low energy region. Energy calibration was done with the internal cosmogenic x-ray peaks: 10.37 keVee of $^{68}$Ge and 8.98 keVee of $^{65}$Zn, and the zero energy defined by the RT events. Analysis procedures follow those with similar detectors in CDEX-1B~\cite{cdex12018}. Basic filtering algorithms are first applied to the $S_{p6,12}$ and $T_p$ pedestals to reject events with anomalous electronic noise profiles. These cuts are energy independent, and the efficiency is measured to be 97.4\% by the survival of RT events, giving rise to a valid data sample of 109.4 days.

\begin{figure}[b] 
\includegraphics[width=0.95\linewidth]{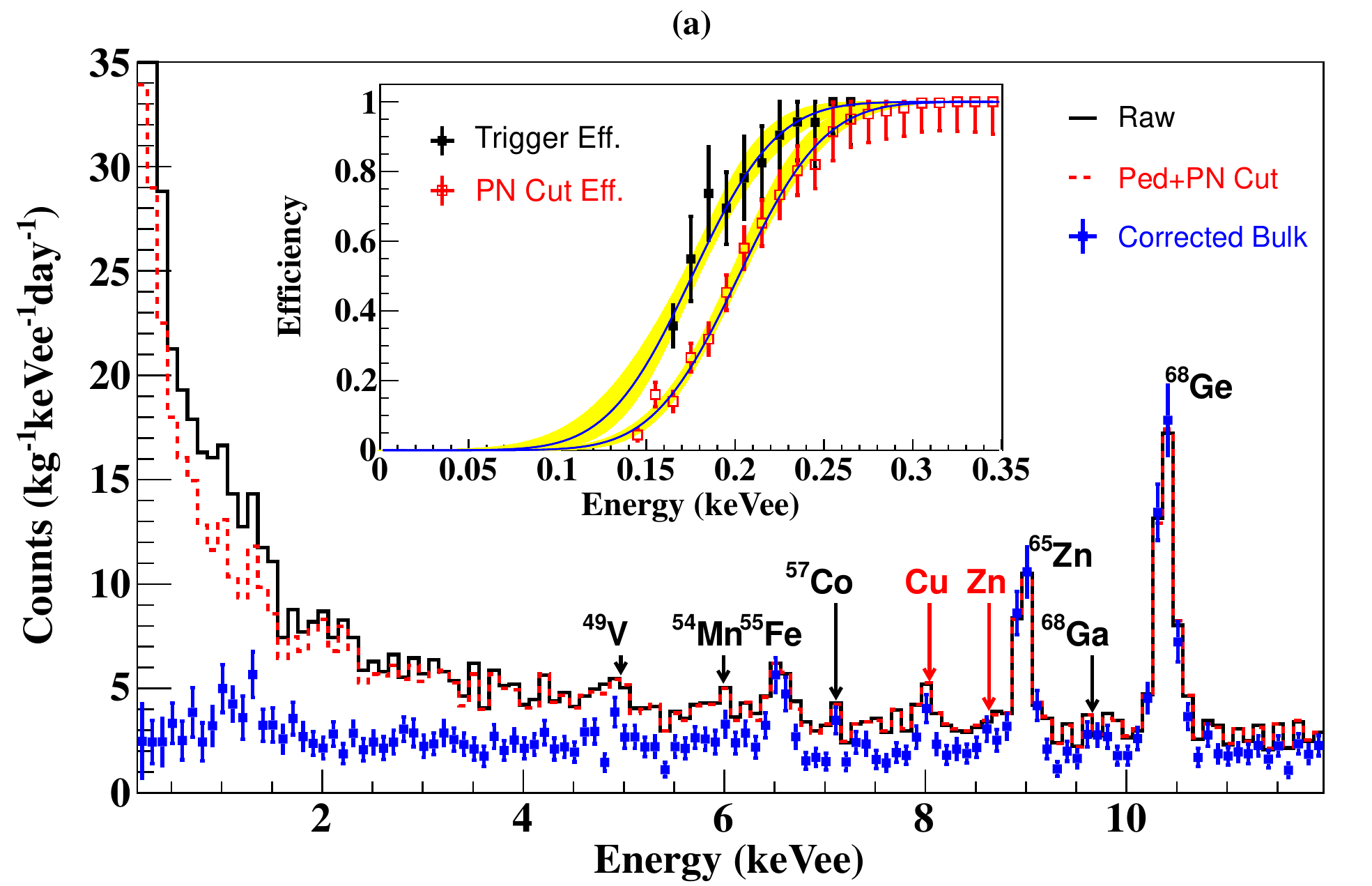}
\includegraphics[width=0.45\linewidth]{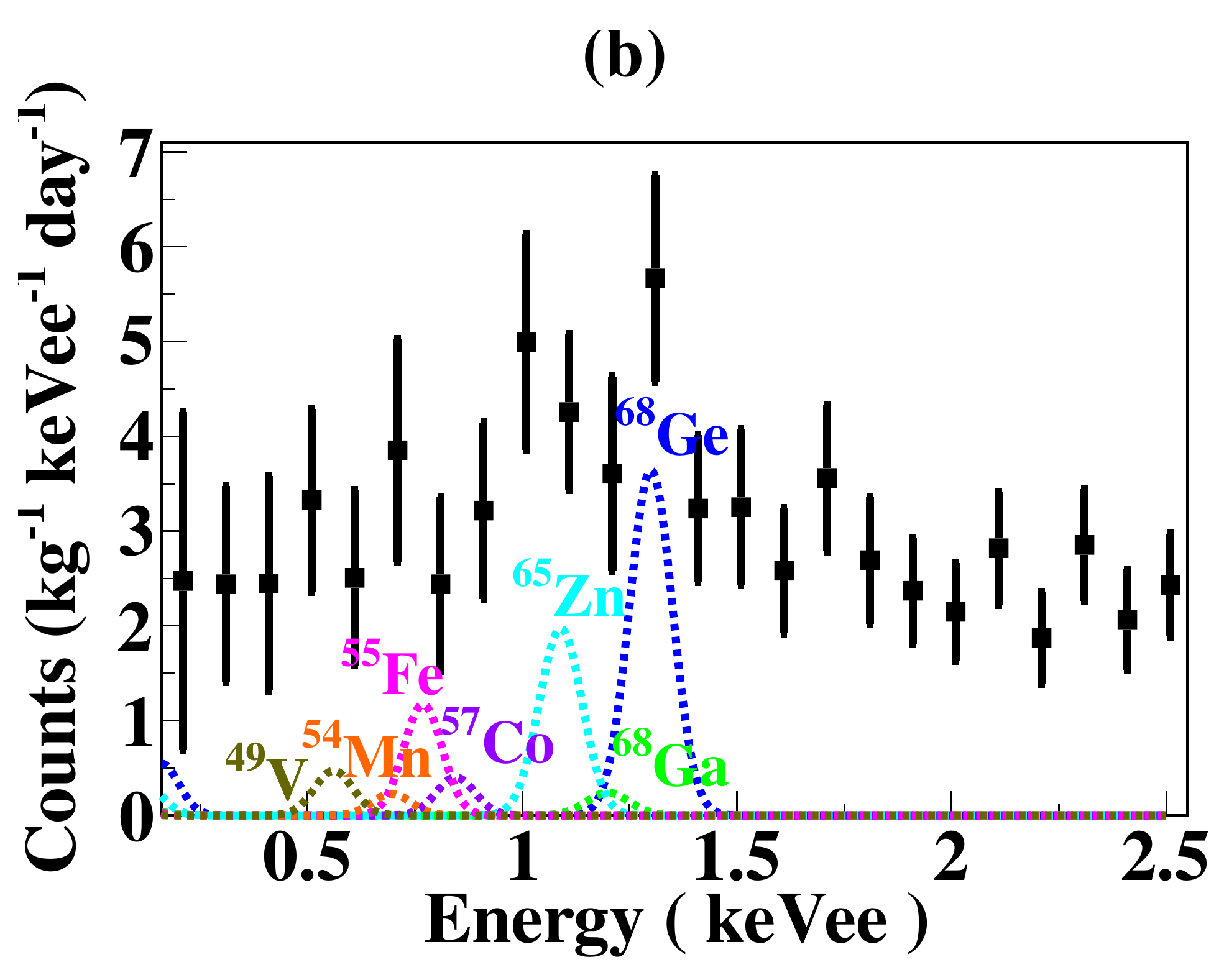}   \includegraphics[width=0.45\linewidth]{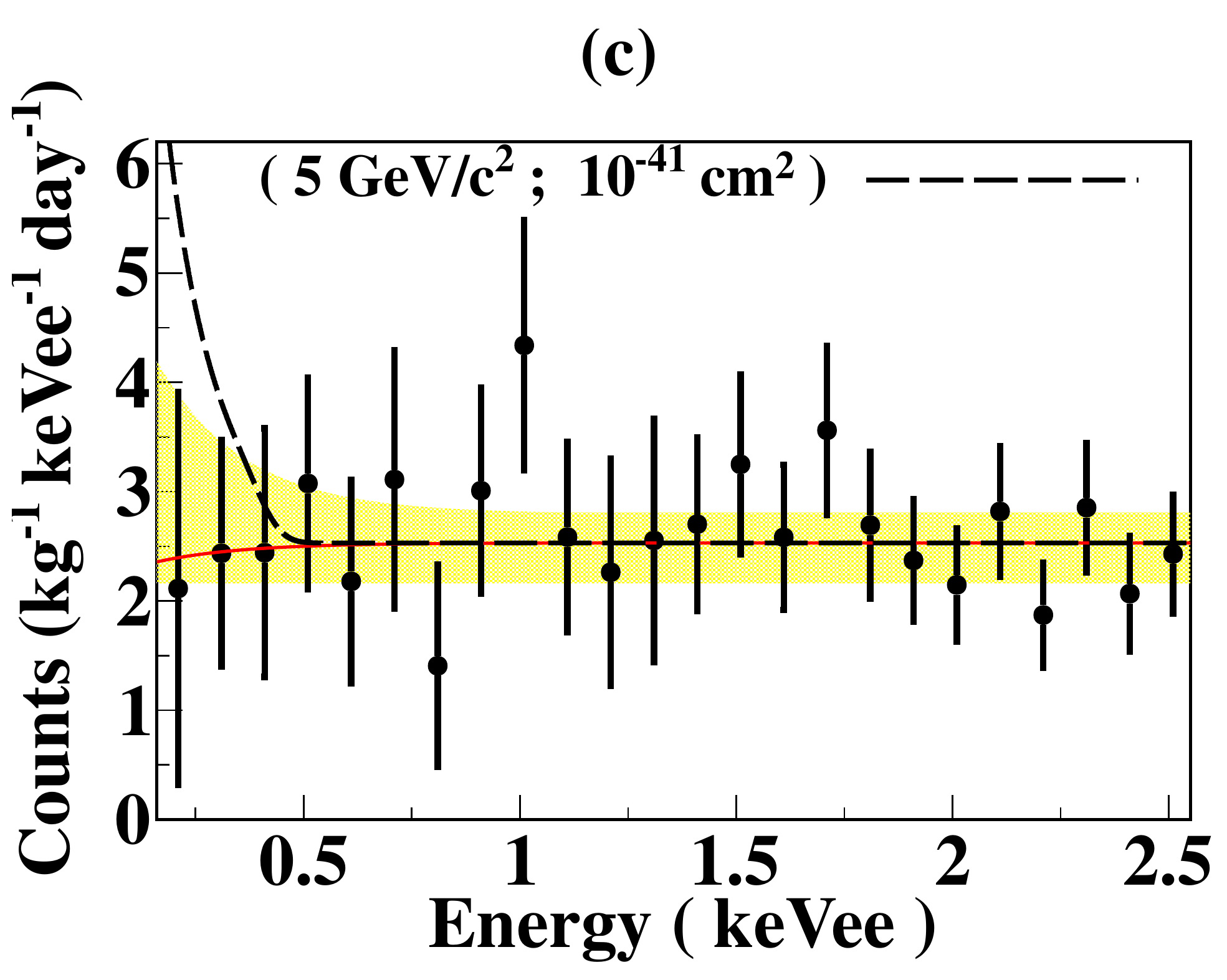}
\caption{
(a) Spectra after different event-selection cuts. The trigger efficiency and PN cut efficiency curves derived from $^{137}$Cs source events and fitted by an error function with a 1$\sigma$ band (yellow shadow) are shown in the inset. (b) $L$-$X$ and $M$-$X$ lines predicted by the $K$-$X$ intensities~\cite{KLratio}. (c) Residual spectrum with the $L$-$X$ and $M$-$X$ contributions subtracted, together with the best-fit spectrum at $m_{\chi}$ = 5 GeV/${c}^2$ (red line), with an uncertainty band (yellow shadow) at the 90\% confidence level. An excluded case at $m_{\chi}$ = 5 GeV/${c}^2$, $\sigma_{\chi N}^{SI} = 10^{-41} \rm{cm}^2$ is superimposed as a black dashed line for illustration.
}
\label{fig::residualspec}
\end{figure}

The second step is a physics-noise event (PN) cut to discriminate the signals from electronic noises near the energy threshold. The PN cut is based on the relationship between the energy and maximum amplitude of $S_{p12}$. The experimental data of a $^{137}$Cs source are used to derive the PN cut and the trigger efficiencies. The efficiency curves with 1$\sigma$ bands are shown in the inset of Fig.~\ref{fig::residualspec}(a).

Events depositing energy in the \emph{n$^{+}$} surface layer generate a slow rising pulse and an incomplete charge collection due to the weak electric field and severe recombination of electron-hole pairs in this region~\cite{LiHB_2014a}. Since C10B-Ge1 and CDEX-1B detectors have the same crystal mass, crystal structure, and fabrication procedure, the same dead layer thickness of 0.88 $\pm$ 0.12 mm \cite{deadlayer} is taken for this analysis~. This gives rise to a fiducial mass of 939 g and accordingly a physics data size of 102.8 kg day. 

\begin{figure}[t] 
\includegraphics[width=0.95\linewidth]{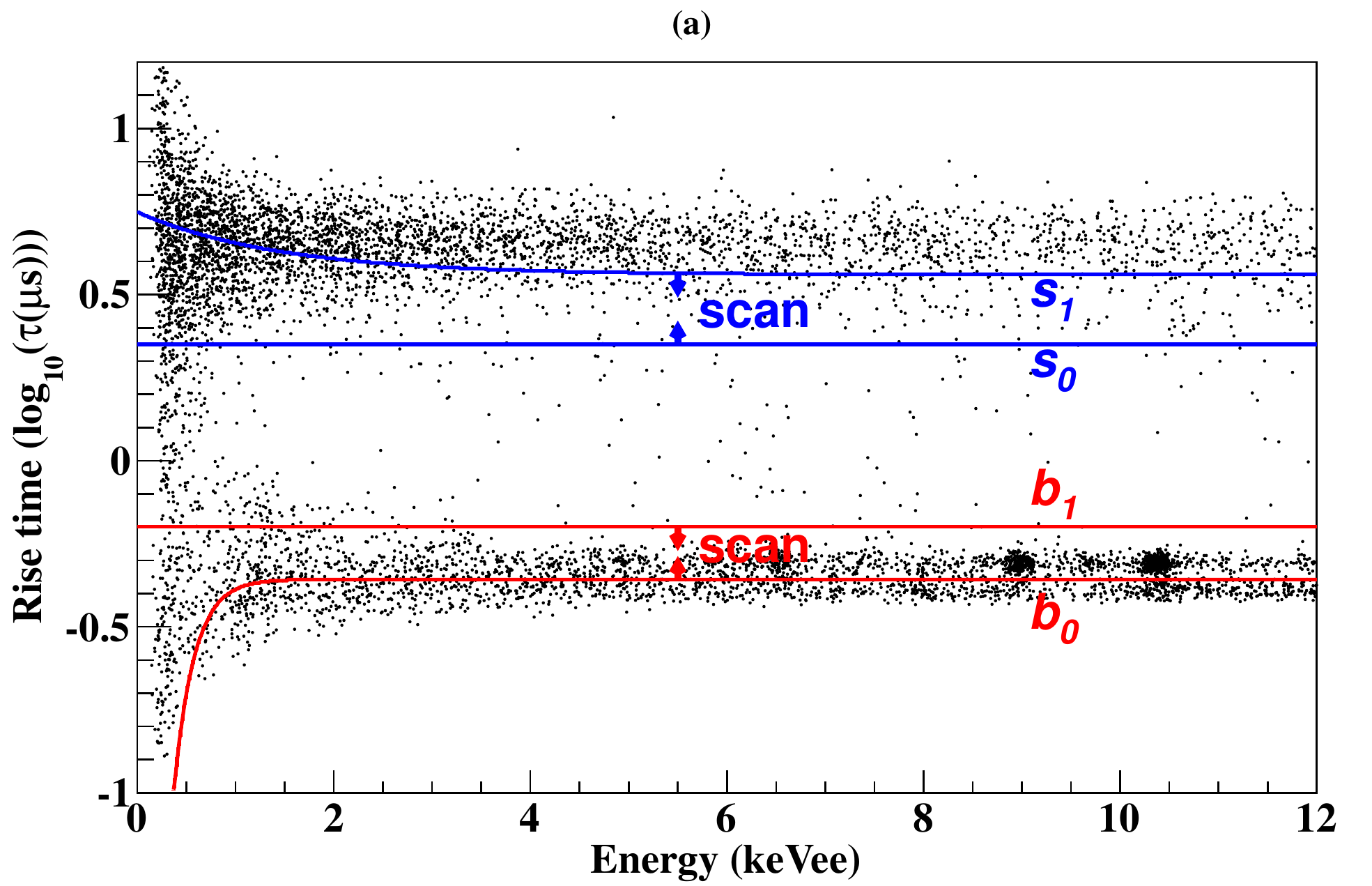}
\includegraphics[width=0.92\linewidth]{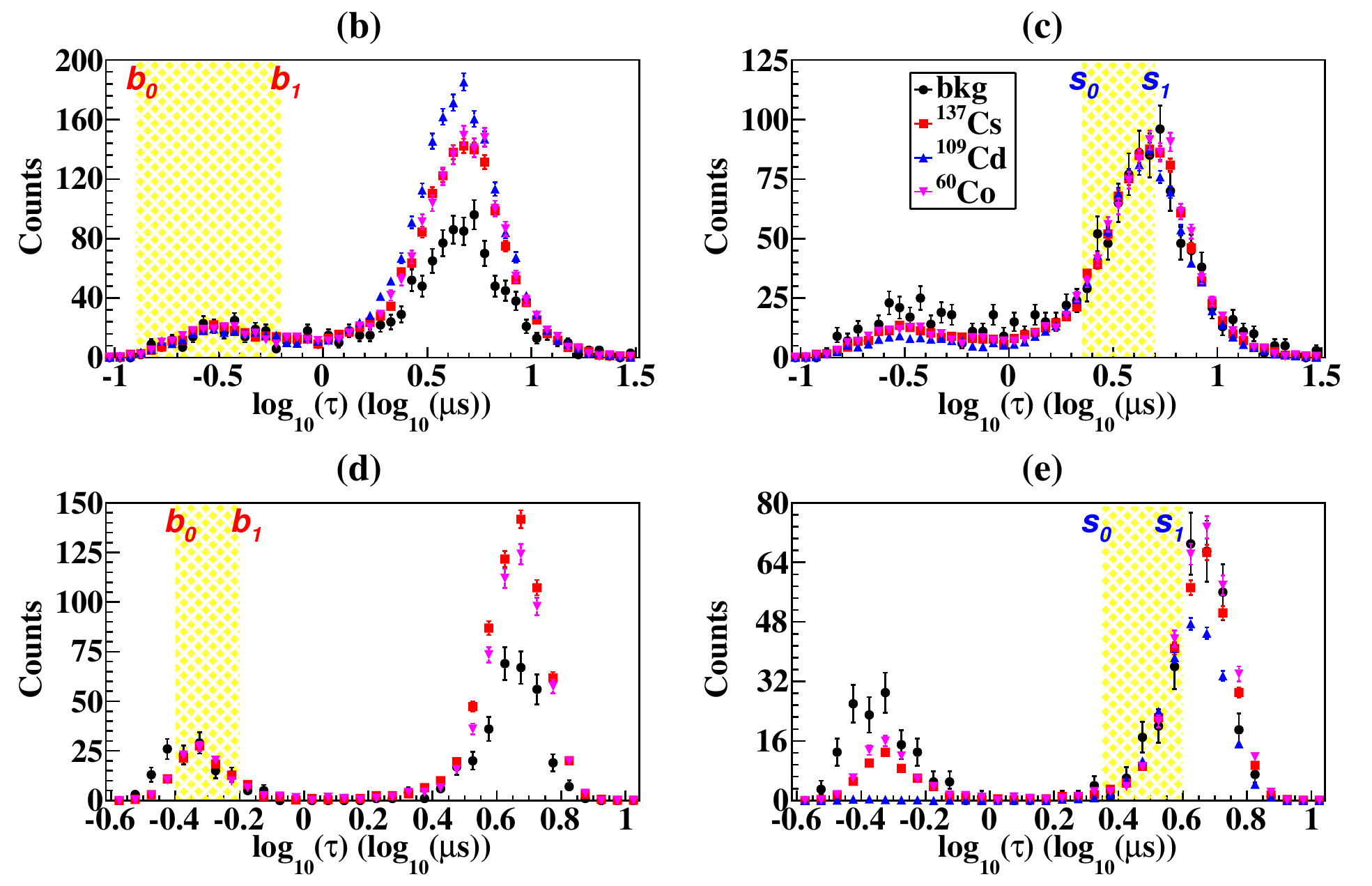}
\caption{
(a) Scatter plot of the rise time [log$_{10}(\tau$)] versus deposited energy of background events. [$b_{0}$, $b_{1}$] and [$s_{0}$, $s_{1}$] are the ``pure" regions we chose to derive the count rates. Extremely-fast and extremely-slow events are with log$_{10}(\tau)<b_0$ and $>s_1$, respectively”. Comparison of the rise-time distribution of various sources and background at typical energies of 0.16-0.66 keVee (b),(c) and 1.66-2.16 keVee (d),(e) with the normalization related to the ``pure" bulk and surface regions (yellow shadow), respectively.
}
\label{fig::bsplot}
\end{figure}

The bulk and surface events (BS) cut is carried out to select bulk events. WIMP candidate events in the bulk of the detector are then separated from the surface events via the rise-time differences of the $T_{p}$ signals. The rise-times ($\tau$) are measured by fitting the $T_{p}$ pulse to a hyperbolic tangent function~\cite{LiHB_2014a,cdex12014,cdex12016,cdex12018}. The log$_{10}(\tau$) distribution versus measured energy of {\it in situ} events is depicted in Fig.~\ref{fig::bsplot}(a), showing a two-band structure of bulk and surface events well separated above 1.5 keVee.  However, at lower energies the bulk and surface events infiltrate into each other, as a result of the electronic noise smearing effect. Multisite events are located off band and of negligible fraction at the keVee-range energy  \cite{soma2016}.

It has been shown that the background and calibration sources data share the common bulk or surface rise-time distribution probability density function (PDF)~\cite{ratiomethod}. The ratio method has been developed accordingly to address the BS discrimination problem in $p$PCGe~\cite{ratiomethod,cdex12018}. In this analysis, the inputs of the ratio method include the background data and three calibration samples ($^{137}$Cs, $^{60}$Co, $^{109}$Cd), while $^{109}$Cd is a pure surface source. Considering that the low-energy gammas from the $^{109}$Cd source can hardly penetrate the \emph{n}$^{+}$ surface layer, their rise-time distribution can describe the surface PDF. Four boundary parameters related to the approximately ``pure" bulk and surface regions are depicted in Fig.~\ref{fig::bsplot}(a). Two outside boundaries [log$_{10}(\tau)$ = $b_0$ and log$_{10}(\tau)$ = $s_{1}$] are derived by fitting the best normalization interval of each energy bin of 500 eVee from 160 eVee on, based on the selection principles of making the statistics as significant as possible while the rise-time distributions of those events remain as consistent as possible. As depicted in Figs.~\ref{fig::bsplot}(b) and \ref{fig::bsplot}(c) and Figs.~\ref{fig::bsplot}(d) and \ref{fig::bsplot}(e), the comparisons of the rise-time distributions of those samples at 0.16-0.66 keVee and 1.66-2.16 keVee demonstrate that they share common rise-time distribution PDFs when normalized to the ``pure" bulk and surface regions. 

\newcommand{\tabincell}[2]{\begin{tabular}{@{}#1@{}}#2\end{tabular}}
\renewcommand\tablename{TABLE}
\renewcommand{\thetable}{\arabic{table}}
\begin{table}
\begin{ruledtabular}
\caption{\label{sys. err} Main contribution to errors of the $B_{r}$ at the threshold bin and a typical high energy bin.}
\centering
\begin{tabular}{lcc}
Energy bin & 0.16-0.26 keVee & 1.96-2.06 keVee \\
\hline
I) Statistic errors  & \hspace{0cm}1.14 & \hspace{0cm}0.50 \\
\hline
II) Systematic errors  \\
   \hspace{0.1cm}(i) Choice of [$b_{0}$, $b_{1}$] & \multirow{2}{*}{1.21} & \multirow{2}{*}{0.10} \\ \hspace{0.5cm}and [$s_{0}$, $s_{1}$]   \\
   \hspace{0.1cm}(ii) Choice of sources  & 0.09 & 0.05 \\
   \hspace{0.1cm}(iii) $\tau$ rebin size  & 0.63 & 0.06 \\
   \hspace{0.1cm}(iv) shift of $\tau$ & 0.06 & 0.01 \\
   Combined  & 1.37 & 0.13 \\
\hline
\tabincell{l}{$B_{r}$ and Errors \\ (kg$^{-1}$keVee$^{-1}$day$^{-1}$)} &  
\tabincell{c}{2.47 $\pm$ 1.14[stat.] \\ \hspace{0.46cm} $\pm$ 1.37[sys.] \\ =2.47 $\pm$ 1.78} & 
\tabincell{c}{2.15 $\pm$ 0.50[stat.] \\  \hspace{0.56cm}$\pm$ 0.13[sys.] \\ =2.15 $\pm$ 0.52}  \\
\end{tabular}
\end{ruledtabular}
\end{table}

\begin{figure*}[t] 
\includegraphics[width=0.5\linewidth]{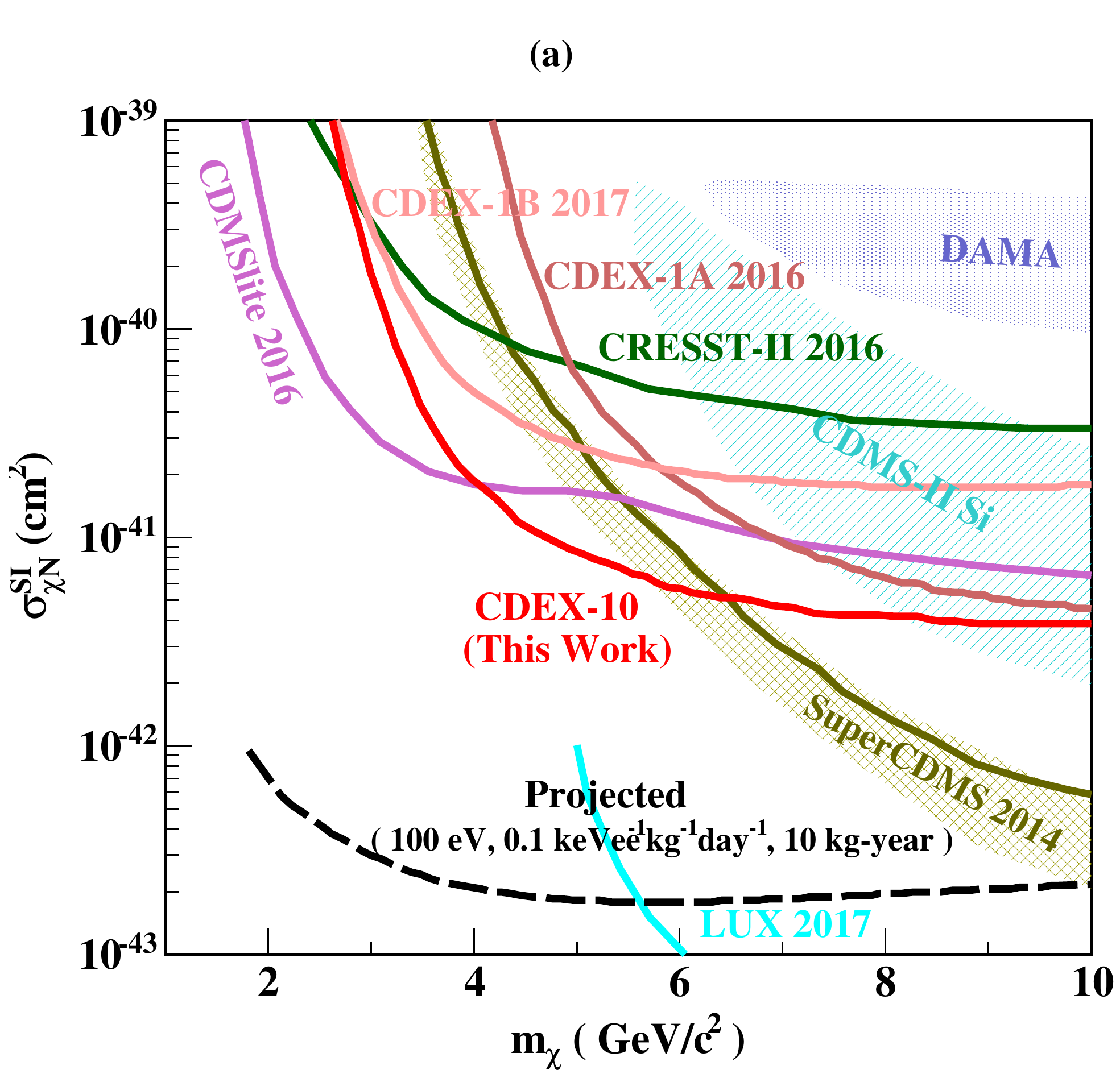}\includegraphics[width=0.5\linewidth]{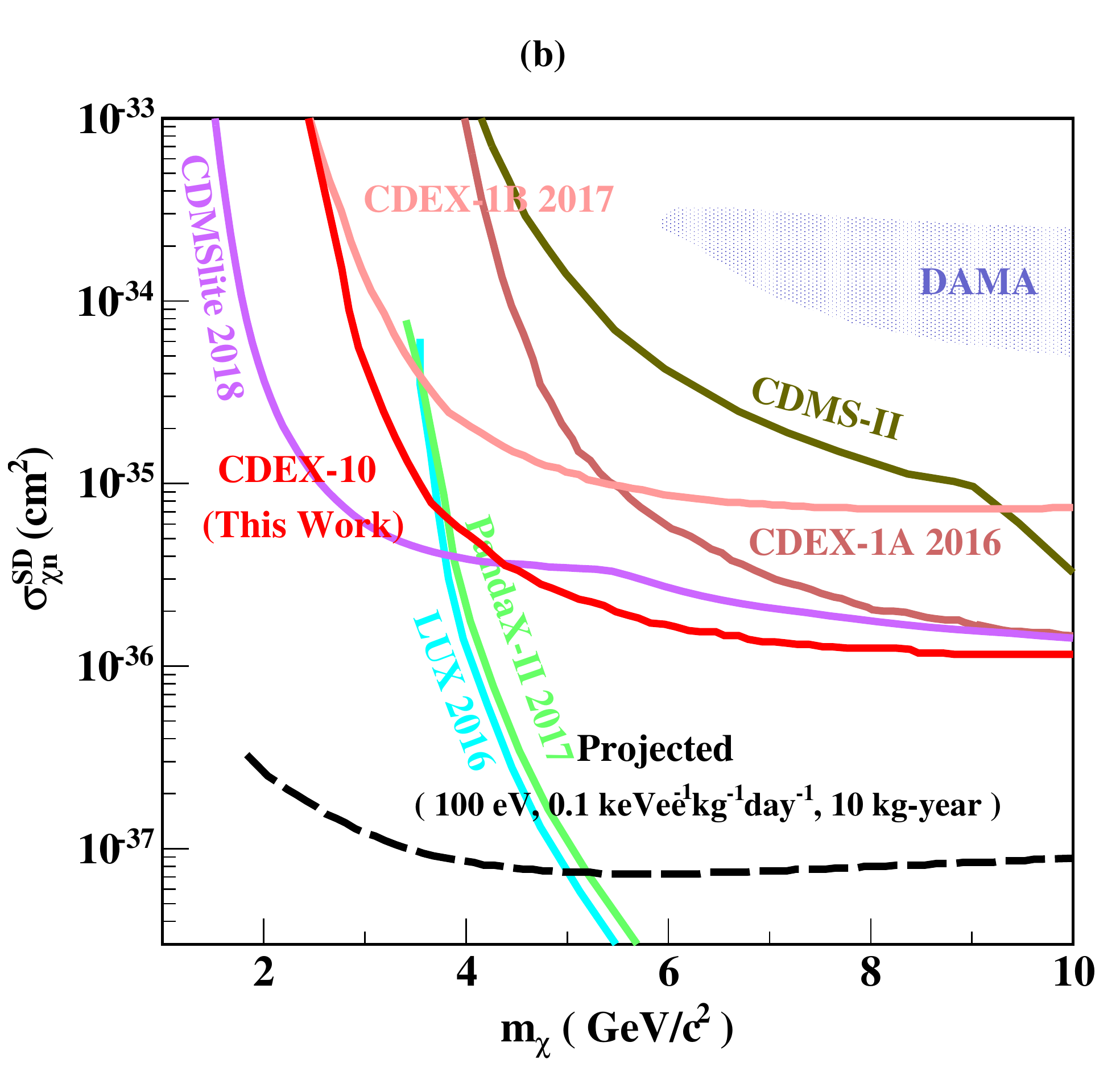}
\caption{
Exclusion plots of (a) SI $\chi$-$N$ coupling and (b) SD $\chi$-neutron coupling  at 90\% C.L., superimposed with results from other benchmark direct search experiments~\cite{cdex12016,cdex12018,cdmslite,supercdms,CRESST-II,DAMA,cdmssi,lux,lux2016,pandax,cdmslite2018}. The best published limits on SI $\chi$-$N$ couplings from the LHC CMS \cite{cms2015,cms3} and ATLAS \cite{atlas1,atlas2,atlas3} experiments are more stringent and beyond the $\csnospin$ scale displayed in (a), though they are extremely model and parameter dependent. New regions on SI for $m_{\chi}$ at 4-5 GeV/$c^2$ are probed and excluded, while liquid xenon experiments~\cite{lux,lux2016,pandax} provide more stringent constraints at $m_{\chi}$ $>$ 5 GeV/$c^2$. The potential reach with target sensitivities of a 100 eVee threshold at 0.1 kg$^{-1}$keVee$^{-1}$day$^{-1}$ background level for 10 kg yr exposure are also superimposed both for SI and SD couplings. 
}
\label{fig::exclusionplot}
\end{figure*}

There are extremely-fast events (EFEs) with a faster rise time in the bulk band due to better rise-time resolution in C10B-Ge1 than CDEX-1A and CDEX-1B~\cite{cdex12016,cdex12018}. It has been verified with simulations using realistic detector electric field that these EFEs mainly originate from the region in the vicinity of the $p^+$ point electrode. An additional convincing evidence is that x rays from Cu are observed only in the EFEs band at 8 keV energy; they can only enter the active area through the passivated surface layer near $p^+$ point. Unfortunately, EFEs can only be distinguished clearly from the bulk band above sub-keVee, while the differentiation is not possible at a low energy region due to the smearing from electronic noise. A cut [log$_{10}$($\tau$) $<$ $b_0$] was used to remove the EFEs, together with an extremely-slow events cut [log$_{10}(\tau) > s_1$]~\cite{ratiomethod} to remove those events which are seriously attenuated by the $n^+$ surface layer. Both kinds of events are included to bulk and surface counts after the $B$ and $S$ correction procedures~\cite{ratiomethod}.

The corrected bulk or surface counts ($B_r$/$S_r$) can be derived by integrating the optimized PDFs which are derived by numerically minimizing the $\chi^2$ of Eq. (7) in Ref.~\cite{ratiomethod}.  The reconstructed $^{137}$Cs and $^{60}$Co spectra are consistent with GEANT4~\cite{Geant4} simulations. The $B_r$ of the background from C10B-Ge1 with the main contributions of errors at the first bin of 0.16-0.26 keVee and a typical high energy of 1.96-2.06 keVee are shown in Table 1. The systematic errors mainly come from the choices of $b_0$, $b_1$, $s_0$, and $s_1$, the errors of which are estimated by varying the more ``pure" bulk and surface regions of Fig.~\ref{fig::bsplot}(a). Further details of the $BS$ analysis and  uncertainties derivations can be found in Ref.~\cite{ratiomethod}.

The spectra after different event-selection cuts are shown in Fig.~\ref{fig::residualspec}(a). The physics analysis threshold is 160 eVee. From the spectra in Fig.~\ref{fig::residualspec}(a), characteristic $K$-shell x ray peaks from internal cosmogenic radionuclides like $^{68,71}$Ge, $^{68}$Ga, $^{65}$Zn, $^{57}$Co, $^{55}$Fe, $^{54}$Mn and $^{49}$V can be identified. In addition, x-ray peaks from Cu and Zn isotopes excited by high energy $\gamma$ rays are observed in the extremely-fast events region of the background spectrum. Their intensities are derived by best fit from the spectrum \cite{cdex12014,cdex12016,cdex12018}. The contributions of $L$- or $M$-shell x-ray peaks are derived from corresponding $K$-shell peaks and subtracted from the $B_{r}$ spectrum, shown in Fig.~\ref{fig::residualspec}(b)~\cite {KLratio}. A minimum-$\chi^2$ analysis~\cite{cdex12014} is applied to the residual spectrum, using two free parameters characterizing the flat background and the possible $\chi$-$N$ SI cross-section ($\csnospin$). The best-fit spectrum at $m_{\chi}$ = 5 GeV/${c}^2$ where $\csnospin = (-0.61 \pm 4.3)\times10^{-42}$ cm$^2$ at $\chi^2/\rm{DOF} = 11.2/22$ ($p$ value = 0.97), is also depicted in Fig.~\ref{fig::residualspec}(c). A standard WIMP galactic halo assumption and conventional astrophysical models~\cite{astropara} are used to describe WIMP-induced interactions, with the local WIMP density of 0.3 GeV/cm$^{3}$, the Maxwellian velocity distribution of $v_0 = 220$ km/s, and the escape velocity of $v_{\rm esc}$ = 544~km/s. The quenching factor in Ge is calculated by the TRIM software package~\cite{trim1,trim2,soma2016,trim3} with a 10\% systematic error adopted for the analysis \cite{cdex12016}.

Upper limits are derived following standard procedures \cite{stat_method,cdex12016}. The exclusion plots of SI and SD at a 90\% confidence level (C.L.) are depicted in Figs.~\ref{fig::exclusionplot}(a) and \ref{fig::exclusionplot}(b), respectively, with several selected benchmark direct search experiments superimposed~\cite{cdex12016,cdex12018,supercdms,cdmslite,CRESST-II,DAMA,cdmssi,lux,lux2016,pandax,cdmslite2018}. The most stringent accelerator bounds on SI from the LHC experiments \cite{cms2015,cms3,atlas1,atlas2,atlas3} are more constraining in SI $-$ with $\csnospin < 10^{-48} ~ {\rm cm^2}$ for $m_{\chi} \sim 5 ~ {\rm GeV}$ $-$ than the  scale displayed in Fig. \ref{fig::exclusionplot}(a). They are, however, extremely sensitive to particle physics models and the choice of parameters. The LHC results are derived with $\chi$-proton cross sections and hence unrelated to the SD constraints on $\chi$-neutron cross sections. This study achieves the lowest threshold and background among the various CDEX data set to date, which brings forth almost an order of magnitude improvement over our previous bounds~\cite{cdex12016,cdex12018}. New regions on SI for $m_{\chi}$ at 4-5 GeV/${c}^{2}$ are probed and excluded. The CDEX-10 detector array will be installed in a new large LN$_2$ cryotank with a volume of about 1700 m$^3$ at Hall-C of CJPL-II \cite{cjpl} by the end of 2018, where shielding from ambient radioactivity is provided by the 6 m-thick LN$_2$. The projected parameter space to be probed with a reduced background comparable to the best achieved in germanium detectors~\cite{MJD} is also shown in Fig.~\ref{fig::exclusionplot}.

This work was supported by the National Key Research and Development Program of China (Grant No. 2017YFA0402201) and the National Natural Science Foundation of China (Grants No.11475092, No. 11475099, No. 11675088, No. 11725522).

H. J. and L.P. J. contributed equally to this work.

\emph{Note added.} $-$  We are aware of  stronger light WIMPs constraints on $\csnospin$ reported in a preprint by the DarkSide-50 experiment \cite{darkside}.

\bibliography{CDEX10.bib}

 \newcommand{\noop}[1]{}
\begin{thebibliography}{39}%
\makeatletter
\providecommand \@ifxundefined [1]{%
 \@ifx{#1\undefined}
}%
\providecommand \@ifnum [1]{%
 \ifnum #1\expandafter \@firstoftwo
 \else \expandafter \@secondoftwo
 \fi
}%
\providecommand \@ifx [1]{%
 \ifx #1\expandafter \@firstoftwo
 \else \expandafter \@secondoftwo
 \fi
}%
\providecommand \natexlab [1]{#1}%
\providecommand \enquote  [1]{``#1''}%
\providecommand \bibnamefont  [1]{#1}%
\providecommand \bibfnamefont [1]{#1}%
\providecommand \citenamefont [1]{#1}%
\providecommand \href@noop [0]{\@secondoftwo}%
\providecommand \href [0]{\begingroup \@sanitize@url \@href}%
\providecommand \@href[1]{\@@startlink{#1}\@@href}%
\providecommand \@@href[1]{\endgroup#1\@@endlink}%
\providecommand \@sanitize@url [0]{\catcode `\\12\catcode `\$12\catcode
  `\&12\catcode `\#12\catcode `\^12\catcode `\_12\catcode `\%12\relax}%
\providecommand \@@startlink[1]{}%
\providecommand \@@endlink[0]{}%
\providecommand \url  [0]{\begingroup\@sanitize@url \@url }%
\providecommand \@url [1]{\endgroup\@href {#1}{\urlprefix }}%
\providecommand \urlprefix  [0]{URL }%
\providecommand \Eprint [0]{\href }%
\providecommand \doibase [0]{http://dx.doi.org/}%
\providecommand \selectlanguage [0]{\@gobble}%
\providecommand \bibinfo  [0]{\@secondoftwo}%
\providecommand \bibfield  [0]{\@secondoftwo}%
\providecommand \translation [1]{[#1]}%
\providecommand \BibitemOpen [0]{}%
\providecommand \bibitemStop [0]{}%
\providecommand \bibitemNoStop [0]{.\EOS\space}%
\providecommand \EOS [0]{\spacefactor3000\relax}%
\providecommand \BibitemShut  [1]{\csname bibitem#1\endcsname}%
\let\auto@bib@innerbib\@empty
\bibitem [{\citenamefont {Patrignani}\ \emph {et~al.}(2016)\citenamefont
  {Patrignani} \emph {et~al.}}]{PDG2017}%
  \BibitemOpen
  \bibfield  {author} {\bibinfo {author} {\bibfnamefont {C.}~\bibnamefont
  {Patrignani}} \emph {et~al.},\ }\href {http://cds.cern.ch/record/2241948}
  {\bibfield  {journal} {\bibinfo  {journal} {Chin. Phys. C}\ }\textbf
  {\bibinfo {volume} {40}},\ \bibinfo {pages} {100001} (\bibinfo {year}
  {2016})}\BibitemShut {NoStop}%
\bibitem [{\citenamefont {Lewin}\ and\ \citenamefont {Smith}(1996)}]{tech}%
  \BibitemOpen
  \bibfield  {author} {\bibinfo {author} {\bibfnamefont {J.}~\bibnamefont
  {Lewin}}\ and\ \bibinfo {author} {\bibfnamefont {P.}~\bibnamefont {Smith}},\
  }\href {\doibase https://doi.org/10.1016/S0927-6505(96)00047-3} {\bibfield
  {journal} {\bibinfo  {journal} {Astropart. Phys.}\ }\textbf {\bibinfo
  {volume} {6}},\ \bibinfo {pages} {87 } (\bibinfo {year} {1996})}\BibitemShut
  {NoStop}%
\bibitem [{\citenamefont {Akerib}\ \emph {et~al.}(2017)\citenamefont {Akerib}
  \emph {et~al.}}]{lux}%
  \BibitemOpen
  \bibfield  {author} {\bibinfo {author} {\bibfnamefont {D.~S.}\ \bibnamefont
  {Akerib}} \emph {et~al.},\ }\href {\doibase 10.1103/PhysRevLett.118.021303}
  {\bibfield  {journal} {\bibinfo  {journal} {Phys. Rev. Lett.}\ }\textbf
  {\bibinfo {volume} {118}},\ \bibinfo {pages} {021303} (\bibinfo {year}
  {2017})}\BibitemShut {NoStop}%
\bibitem [{\citenamefont {Cui}\ \emph {et~al.}(2017)\citenamefont {Cui} \emph
  {et~al.}}]{pandax}%
  \BibitemOpen
  \bibfield  {author} {\bibinfo {author} {\bibfnamefont {X.}~\bibnamefont
  {Cui}} \emph {et~al.},\ }\href {\doibase 10.1103/PhysRevLett.119.181302}
  {\bibfield  {journal} {\bibinfo  {journal} {Phys. Rev. Lett.}\ }\textbf
  {\bibinfo {volume} {119}},\ \bibinfo {pages} {181302} (\bibinfo {year}
  {2017})}\BibitemShut {NoStop}%
\bibitem [{\citenamefont {Aprile}\ \emph {et~al.}(2017)\citenamefont {Aprile}
  \emph {et~al.}}]{xenon}%
  \BibitemOpen
  \bibfield  {author} {\bibinfo {author} {\bibfnamefont {E.}~\bibnamefont
  {Aprile}} \emph {et~al.},\ }\href {\doibase 10.1103/PhysRevLett.119.181301}
  {\bibfield  {journal} {\bibinfo  {journal} {Phys. Rev. Lett.}\ }\textbf
  {\bibinfo {volume} {119}},\ \bibinfo {pages} {181301} (\bibinfo {year}
  {2017})}\BibitemShut {NoStop}%
\bibitem [{\citenamefont {Aalseth}\ \emph
  {et~al.}(2013{\natexlab{a}})\citenamefont {Aalseth} \emph {et~al.}}]{cogent}%
  \BibitemOpen
  \bibfield  {author} {\bibinfo {author} {\bibfnamefont {C.~E.}\ \bibnamefont
  {Aalseth}} \emph {et~al.},\ }\href {\doibase 10.1103/PhysRevD.88.012002}
  {\bibfield  {journal} {\bibinfo  {journal} {Phys. Rev. D}\ }\textbf {\bibinfo
  {volume} {88}},\ \bibinfo {pages} {012002} (\bibinfo {year}
  {2013}{\natexlab{a}})}\BibitemShut {NoStop}%
\bibitem [{\citenamefont {Liu}\ \emph {et~al.}(2014)\citenamefont {Liu} \emph
  {et~al.}}]{cdex0}%
  \BibitemOpen
  \bibfield  {author} {\bibinfo {author} {\bibfnamefont {S.~K.}\ \bibnamefont
  {Liu}} \emph {et~al.},\ }\href {\doibase 10.1103/PhysRevD.90.032003}
  {\bibfield  {journal} {\bibinfo  {journal} {Phys. Rev. D}\ }\textbf {\bibinfo
  {volume} {90}},\ \bibinfo {pages} {032003} (\bibinfo {year}
  {2014})}\BibitemShut {NoStop}%
\bibitem [{\citenamefont {Yue}\ \emph {et~al.}(2014)\citenamefont {Yue} \emph
  {et~al.}}]{cdex12014}%
  \BibitemOpen
  \bibfield  {author} {\bibinfo {author} {\bibfnamefont {Q.}~\bibnamefont
  {Yue}} \emph {et~al.},\ }\href {\doibase 10.1103/PhysRevD.90.091701}
  {\bibfield  {journal} {\bibinfo  {journal} {Phys. Rev. D}\ }\textbf {\bibinfo
  {volume} {90}},\ \bibinfo {pages} {091701} (\bibinfo {year}
  {2014})}\BibitemShut {NoStop}%
\bibitem [{\citenamefont {Zhao}\ \emph {et~al.}(2016)\citenamefont {Zhao} \emph
  {et~al.}}]{cdex12016}%
  \BibitemOpen
  \bibfield  {author} {\bibinfo {author} {\bibfnamefont {W.}~\bibnamefont
  {Zhao}} \emph {et~al.},\ }\href {\doibase 10.1103/PhysRevD.93.092003}
  {\bibfield  {journal} {\bibinfo  {journal} {Phys. Rev. D}\ }\textbf {\bibinfo
  {volume} {93}},\ \bibinfo {pages} {092003} (\bibinfo {year}
  {2016})}\BibitemShut {NoStop}%
\bibitem [{\citenamefont {Yang}\ \emph
  {et~al.}(2018{\natexlab{a}})\citenamefont {Yang} \emph {et~al.}}]{cdex12018}%
  \BibitemOpen
  \bibfield  {author} {\bibinfo {author} {\bibfnamefont {L.~T.}\ \bibnamefont
  {Yang}} \emph {et~al.},\ }\href {\doibase 10.1088/1674-1137/42/2/023002}
  {\bibfield  {journal} {\bibinfo  {journal} {Chin. Phys. C}\ }\textbf
  {\bibinfo {volume} {42}},\ \bibinfo {eid} {23002} (\bibinfo {year}
  {2018}{\natexlab{a}})}\BibitemShut {NoStop}%
\bibitem [{\citenamefont {Agnese}\ \emph {et~al.}(2014)\citenamefont {Agnese}
  \emph {et~al.}}]{supercdms}%
  \BibitemOpen
  \bibfield  {author} {\bibinfo {author} {\bibfnamefont {R.}~\bibnamefont
  {Agnese}} \emph {et~al.},\ }\href {\doibase 10.1103/PhysRevLett.112.241302}
  {\bibfield  {journal} {\bibinfo  {journal} {Phys. Rev. Lett.}\ }\textbf
  {\bibinfo {volume} {112}},\ \bibinfo {pages} {241302} (\bibinfo {year}
  {2014})}\BibitemShut {NoStop}%
\bibitem [{\citenamefont {Agnese}\ \emph {et~al.}(2016)\citenamefont {Agnese}
  \emph {et~al.}}]{cdmslite}%
  \BibitemOpen
  \bibfield  {author} {\bibinfo {author} {\bibfnamefont {R.}~\bibnamefont
  {Agnese}} \emph {et~al.},\ }\href {\doibase 10.1103/PhysRevLett.116.071301}
  {\bibfield  {journal} {\bibinfo  {journal} {Phys. Rev. Lett.}\ }\textbf
  {\bibinfo {volume} {116}},\ \bibinfo {pages} {071301} (\bibinfo {year}
  {2016})}\BibitemShut {NoStop}%
\bibitem [{\citenamefont {Angloher}\ \emph {et~al.}(2016)\citenamefont
  {Angloher} \emph {et~al.}}]{CRESST-II}%
  \BibitemOpen
  \bibfield  {author} {\bibinfo {author} {\bibfnamefont {G.}~\bibnamefont
  {Angloher}} \emph {et~al.},\ }\href {\doibase 10.1140/epjc/s10052-016-3877-3}
  {\bibfield  {journal} {\bibinfo  {journal} {Eur. Phys. J. C}\ }\textbf
  {\bibinfo {volume} {76}},\ \bibinfo {pages} {25} (\bibinfo {year}
  {2016})}\BibitemShut {NoStop}%
\bibitem [{\citenamefont {Cheng}\ \emph {et~al.}(2017)\citenamefont {Cheng}
  \emph {et~al.}}]{cjpl}%
  \BibitemOpen
  \bibfield  {author} {\bibinfo {author} {\bibfnamefont {J.~P.}\ \bibnamefont
  {Cheng}} \emph {et~al.},\ }\href {\doibase
  10.1146/annurev-nucl-102115-044842} {\bibfield  {journal} {\bibinfo
  {journal} {Annu. Rev. Nucl. Part. Sci.}\ }\textbf {\bibinfo {volume} {67}},\
  \bibinfo {pages} {231} (\bibinfo {year} {2017})}\BibitemShut {NoStop}%
\bibitem [{\citenamefont {Klapdor-Kleingrothaus}\ \emph
  {et~al.}(1998)\citenamefont {Klapdor-Kleingrothaus}, \citenamefont
  {Hellmig},\ and\ \citenamefont {Hirsch}}]{GENIUS}%
  \BibitemOpen
  \bibfield  {author} {\bibinfo {author} {\bibfnamefont {H.~V.}\ \bibnamefont
  {Klapdor-Kleingrothaus}}, \bibinfo {author} {\bibfnamefont {J.}~\bibnamefont
  {Hellmig}}, \ and\ \bibinfo {author} {\bibfnamefont {M.}~\bibnamefont
  {Hirsch}},\ }\href@noop {} {\bibfield  {journal} {\bibinfo  {journal} {J.
  Phys. G}\ }\textbf {\bibinfo {volume} {24}},\ \bibinfo {pages} {483}
  (\bibinfo {year} {1998})}\BibitemShut {NoStop}%
\bibitem [{\citenamefont {Ackermann}\ \emph {et~al.}(2013)\citenamefont
  {Ackermann} \emph {et~al.}}]{gerda}%
  \BibitemOpen
  \bibfield  {author} {\bibinfo {author} {\bibfnamefont {K.~H.}\ \bibnamefont
  {Ackermann}} \emph {et~al.},\ }\href {\doibase
  10.1140/epjc/s10052-013-2330-0} {\bibfield  {journal} {\bibinfo  {journal}
  {Eur. Phys. J. C}\ }\textbf {\bibinfo {volume} {73}},\ \bibinfo {pages}
  {2330} (\bibinfo {year} {2013})}\BibitemShut {NoStop}%
\bibitem [{\citenamefont {Abgrall}\ \emph
  {et~al.}(2017{\natexlab{a}})\citenamefont {Abgrall} \emph {et~al.}}]{LEGEND}%
  \BibitemOpen
  \bibfield  {author} {\bibinfo {author} {\bibfnamefont {N.}~\bibnamefont
  {Abgrall}} \emph {et~al.},\ }\href {\doibase 10.1063/1.5007652} {\bibfield
  {journal} {\bibinfo  {journal} {AIP Conf. Proc.}\ }\textbf {\bibinfo {volume}
  {1894}},\ \bibinfo {pages} {020027} (\bibinfo {year}
  {2017}{\natexlab{a}})}\BibitemShut {NoStop}%
\bibitem [{\citenamefont {Li}\ \emph {et~al.}(2014)\citenamefont {Li} \emph
  {et~al.}}]{LiHB_2014a}%
  \BibitemOpen
  \bibfield  {author} {\bibinfo {author} {\bibfnamefont {H.~B.}\ \bibnamefont
  {Li}} \emph {et~al.},\ }\href@noop {} {\bibfield  {journal} {\bibinfo
  {journal} {Astropart. Phys.}\ }\textbf {\bibinfo {volume} {56}},\ \bibinfo
  {pages} {1} (\bibinfo {year} {2014})}\BibitemShut {NoStop}%
\bibitem [{\citenamefont {Ma}\ \emph {et~al.}(2017)\citenamefont {Ma} \emph
  {et~al.}}]{deadlayer}%
  \BibitemOpen
  \bibfield  {author} {\bibinfo {author} {\bibfnamefont {J.~L.}\ \bibnamefont
  {Ma}} \emph {et~al.},\ }\href {\doibase
  https://doi.org/10.1016/j.apradiso.2017.05.023} {\bibfield  {journal}
  {\bibinfo  {journal} {Appl. Radiat. Isot.}\ }\textbf {\bibinfo {volume}
  {127}},\ \bibinfo {pages} {130 } (\bibinfo {year} {2017})}\BibitemShut
  {NoStop}%
\bibitem [{\citenamefont {Soma}\ \emph {et~al.}(2016)\citenamefont {Soma} \emph
  {et~al.}}]{soma2016}%
  \BibitemOpen
  \bibfield  {author} {\bibinfo {author} {\bibfnamefont {A.~K.}\ \bibnamefont
  {Soma}} \emph {et~al.},\ }\href {\doibase
  https://doi.org/10.1016/j.nima.2016.08.044} {\bibfield  {journal} {\bibinfo
  {journal} {Nucl. Instrum. Meth. A}\ }\textbf {\bibinfo {volume} {836}},\
  \bibinfo {pages} {67 } (\bibinfo {year} {2016})}\BibitemShut {NoStop}%
\bibitem [{\citenamefont {Yang}\ \emph
  {et~al.}(2018{\natexlab{b}})\citenamefont {Yang} \emph
  {et~al.}}]{ratiomethod}%
  \BibitemOpen
  \bibfield  {author} {\bibinfo {author} {\bibfnamefont {L.~T.}\ \bibnamefont
  {Yang}} \emph {et~al.},\ }\href {\doibase
  https://doi.org/10.1016/j.nima.2017.12.078} {\bibfield  {journal} {\bibinfo
  {journal} {Nucl. Instrum. Meth. A}\ }\textbf {\bibinfo {volume} {886}},\
  \bibinfo {pages} {13 } (\bibinfo {year} {2018}{\natexlab{b}})}\BibitemShut
  {NoStop}%
\bibitem [{\citenamefont {Bahcall}(1963)}]{KLratio}%
  \BibitemOpen
  \bibfield  {author} {\bibinfo {author} {\bibfnamefont {J.~N.}\ \bibnamefont
  {Bahcall}},\ }\href {\doibase 10.1103/PhysRev.132.362} {\bibfield  {journal}
  {\bibinfo  {journal} {Phys. Rev.}\ }\textbf {\bibinfo {volume} {132}},\
  \bibinfo {pages} {362} (\bibinfo {year} {1963})}\BibitemShut {NoStop}%
\bibitem [{\citenamefont {Belli}\ \emph {et~al.}(2011)\citenamefont {Belli}
  \emph {et~al.}}]{DAMA}%
  \BibitemOpen
  \bibfield  {author} {\bibinfo {author} {\bibfnamefont {P.}~\bibnamefont
  {Belli}} \emph {et~al.},\ }\href {\doibase 10.1103/PhysRevD.84.055014}
  {\bibfield  {journal} {\bibinfo  {journal} {Phys. Rev. D}\ }\textbf {\bibinfo
  {volume} {84}},\ \bibinfo {pages} {055014} (\bibinfo {year}
  {2011})}\BibitemShut {NoStop}%
\bibitem [{\citenamefont {Agnese}\ \emph {et~al.}(2013)\citenamefont {Agnese}
  \emph {et~al.}}]{cdmssi}%
  \BibitemOpen
  \bibfield  {author} {\bibinfo {author} {\bibfnamefont {R.}~\bibnamefont
  {Agnese}} \emph {et~al.},\ }\href {\doibase 10.1103/PhysRevLett.111.251301}
  {\bibfield  {journal} {\bibinfo  {journal} {Phys. Rev. Lett.}\ }\textbf
  {\bibinfo {volume} {111}},\ \bibinfo {pages} {251301} (\bibinfo {year}
  {2013})}\BibitemShut {NoStop}%
\bibitem [{\citenamefont {Akerib}\ \emph {et~al.}(2016)\citenamefont {Akerib}
  \emph {et~al.}}]{lux2016}%
  \BibitemOpen
  \bibfield  {author} {\bibinfo {author} {\bibfnamefont {D.~S.}\ \bibnamefont
  {Akerib}} \emph {et~al.},\ }\href {\doibase 10.1103/PhysRevLett.116.161302}
  {\bibfield  {journal} {\bibinfo  {journal} {Phys. Rev. Lett.}\ }\textbf
  {\bibinfo {volume} {116}},\ \bibinfo {pages} {161302} (\bibinfo {year}
  {2016})}\BibitemShut {NoStop}%
\bibitem [{\citenamefont {Agnese}\ \emph {et~al.}(2018)\citenamefont {Agnese}
  \emph {et~al.}}]{cdmslite2018}%
  \BibitemOpen
  \bibfield  {author} {\bibinfo {author} {\bibfnamefont {R.}~\bibnamefont
  {Agnese}} \emph {et~al.},\ }\href {\doibase 10.1103/PhysRevD.97.022002}
  {\bibfield  {journal} {\bibinfo  {journal} {Phys. Rev. D}\ }\textbf {\bibinfo
  {volume} {97}},\ \bibinfo {pages} {022002} (\bibinfo {year}
  {2018})}\BibitemShut {NoStop}%
\bibitem [{\citenamefont {Khachatryan}\ \emph {et~al.}(2015)\citenamefont
  {Khachatryan} \emph {et~al.}}]{cms2015}%
  \BibitemOpen
  \bibfield  {author} {\bibinfo {author} {\bibfnamefont {V.}~\bibnamefont
  {Khachatryan}} \emph {et~al.},\ }\href {\doibase
  10.1140/epjc/s10052-015-3451-4} {\bibfield  {journal} {\bibinfo  {journal}
  {Eur. Phys. J. C}\ }\textbf {\bibinfo {volume} {75}},\ \bibinfo {pages} {235}
  (\bibinfo {year} {2015})}\BibitemShut {NoStop}%
\bibitem [{\citenamefont {Khachatryan}\ \emph {et~al.}(2017)\citenamefont
  {Khachatryan} \emph {et~al.}}]{cms3}%
  \BibitemOpen
  \bibfield  {author} {\bibinfo {author} {\bibfnamefont {V.}~\bibnamefont
  {Khachatryan}} \emph {et~al.},\ }\href {\doibase 10.1007/JHEP02(2017)135}
  {\bibfield  {journal} {\bibinfo  {journal} {Journal of High Energy Physics}\
  }\textbf {\bibinfo {volume} {2017}},\ \bibinfo {pages} {135} (\bibinfo {year}
  {2017})}\BibitemShut {NoStop}%
\bibitem [{\citenamefont {Aad}\ \emph {et~al.}(2015{\natexlab{a}})\citenamefont
  {Aad} \emph {et~al.}}]{atlas1}%
  \BibitemOpen
  \bibfield  {author} {\bibinfo {author} {\bibfnamefont {G.}~\bibnamefont
  {Aad}} \emph {et~al.},\ }\href {\doibase 10.1007/JHEP11(2015)206} {\bibfield
  {journal} {\bibinfo  {journal} {J. High Energy Phys.}\ }\textbf {\bibinfo
  {volume} {2015}},\ \bibinfo {pages} {206} (\bibinfo {year}
  {2015}{\natexlab{a}})}\BibitemShut {NoStop}%
\bibitem [{\citenamefont {Aad}\ \emph {et~al.}(2015{\natexlab{b}})\citenamefont
  {Aad} \emph {et~al.}}]{atlas2}%
  \BibitemOpen
  \bibfield  {author} {\bibinfo {author} {\bibfnamefont {G.}~\bibnamefont
  {Aad}} \emph {et~al.},\ }\href {\doibase 10.1140/epjc/s10052-015-3517-3}
  {\bibfield  {journal} {\bibinfo  {journal} {Eur. Phys. J. C}\ }\textbf
  {\bibinfo {volume} {75}},\ \bibinfo {pages} {299} (\bibinfo {year}
  {2015}{\natexlab{b}})}\BibitemShut {NoStop}%
\bibitem [{\citenamefont {Hoferichter}\ \emph {et~al.}(2017)\citenamefont
  {Hoferichter} \emph {et~al.}}]{atlas3}%
  \BibitemOpen
  \bibfield  {author} {\bibinfo {author} {\bibfnamefont {M.}~\bibnamefont
  {Hoferichter}} \emph {et~al.},\ }\href {\doibase
  10.1103/PhysRevLett.119.181803} {\bibfield  {journal} {\bibinfo  {journal}
  {Phys. Rev. Lett.}\ }\textbf {\bibinfo {volume} {119}},\ \bibinfo {pages}
  {181803} (\bibinfo {year} {2017})}\BibitemShut {NoStop}%
\bibitem [{\citenamefont {Agostinelli}\ \emph {et~al.}(2003)\citenamefont
  {Agostinelli} \emph {et~al.}}]{Geant4}%
  \BibitemOpen
  \bibfield  {author} {\bibinfo {author} {\bibfnamefont {S.}~\bibnamefont
  {Agostinelli}} \emph {et~al.},\ }\href {\doibase
  https://doi.org/10.1016/S0168-9002(03)01368-8} {\bibfield  {journal}
  {\bibinfo  {journal} {Nucl. Instrum. Meth. A}\ }\textbf {\bibinfo {volume}
  {506}},\ \bibinfo {pages} {250 } (\bibinfo {year} {2003})}\BibitemShut
  {NoStop}%
\bibitem [{\citenamefont {Aalseth}\ \emph
  {et~al.}(2013{\natexlab{b}})\citenamefont {Aalseth} \emph
  {et~al.}}]{astropara}%
  \BibitemOpen
  \bibfield  {author} {\bibinfo {author} {\bibfnamefont {C.~E.}\ \bibnamefont
  {Aalseth}} \emph {et~al.},\ }\href {\doibase 10.1103/PhysRevD.88.012002}
  {\bibfield  {journal} {\bibinfo  {journal} {Phys. Rev. D}\ }\textbf {\bibinfo
  {volume} {88}},\ \bibinfo {pages} {012002} (\bibinfo {year}
  {2013}{\natexlab{b}})}\BibitemShut {NoStop}%
\bibitem [{\citenamefont {Ziegler}(2004)}]{trim1}%
  \BibitemOpen
  \bibfield  {author} {\bibinfo {author} {\bibfnamefont {J.~F.}\ \bibnamefont
  {Ziegler}},\ }\href {\doibase https://doi.org/10.1016/j.nimb.2004.01.208}
  {\bibfield  {journal} {\bibinfo  {journal} {Nucl. Instrum. Meth. B}\ }\textbf
  {\bibinfo {volume} {219-220}},\ \bibinfo {pages} {1027 } (\bibinfo {year}
  {2004})}\BibitemShut {NoStop}%
\bibitem [{\citenamefont {Lin}\ \emph {et~al.}(2009)\citenamefont {Lin} \emph
  {et~al.}}]{trim2}%
  \BibitemOpen
  \bibfield  {author} {\bibinfo {author} {\bibfnamefont {S.~T.}\ \bibnamefont
  {Lin}} \emph {et~al.},\ }\href {\doibase 10.1103/PhysRevD.79.061101}
  {\bibfield  {journal} {\bibinfo  {journal} {Phys. Rev. D}\ }\textbf {\bibinfo
  {volume} {79}},\ \bibinfo {pages} {061101} (\bibinfo {year}
  {2009})}\BibitemShut {NoStop}%
\bibitem [{\citenamefont {Scholz}\ \emph {et~al.}(2016)\citenamefont {Scholz}
  \emph {et~al.}}]{trim3}%
  \BibitemOpen
  \bibfield  {author} {\bibinfo {author} {\bibfnamefont {B.~J.}\ \bibnamefont
  {Scholz}} \emph {et~al.},\ }\href {\doibase 10.1103/PhysRevD.94.122003}
  {\bibfield  {journal} {\bibinfo  {journal} {Phys. Rev. D}\ }\textbf {\bibinfo
  {volume} {94}},\ \bibinfo {pages} {122003} (\bibinfo {year}
  {2016})}\BibitemShut {NoStop}%
\bibitem [{\citenamefont {Feldman}\ and\ \citenamefont
  {Cousins}(1998)}]{stat_method}%
  \BibitemOpen
  \bibfield  {author} {\bibinfo {author} {\bibfnamefont {G.~J.}\ \bibnamefont
  {Feldman}}\ and\ \bibinfo {author} {\bibfnamefont {R.~D.}\ \bibnamefont
  {Cousins}},\ }\href {\doibase 10.1103/PhysRevD.57.3873} {\bibfield  {journal}
  {\bibinfo  {journal} {Phys. Rev. D}\ }\textbf {\bibinfo {volume} {57}},\
  \bibinfo {pages} {3873} (\bibinfo {year} {1998})}\BibitemShut {NoStop}%
\bibitem [{\citenamefont {Abgrall}\ \emph
  {et~al.}(2017{\natexlab{b}})\citenamefont {Abgrall} \emph {et~al.}}]{MJD}%
  \BibitemOpen
  \bibfield  {author} {\bibinfo {author} {\bibfnamefont {N.}~\bibnamefont
  {Abgrall}} \emph {et~al.},\ }\href {\doibase 10.1103/PhysRevLett.118.161801}
  {\bibfield  {journal} {\bibinfo  {journal} {Phys. Rev. Lett.}\ }\textbf
  {\bibinfo {volume} {118}},\ \bibinfo {pages} {161801} (\bibinfo {year}
  {2017}{\natexlab{b}})}\BibitemShut {NoStop}%
\bibitem [{\citenamefont {Agnes}\ \emph {et~al.}()\citenamefont {Agnes} \emph
  {et~al.}}]{darkside}%
  \BibitemOpen
  \bibfield  {author} {\bibinfo {author} {\bibfnamefont {P.}~\bibnamefont
  {Agnes}} \emph {et~al.},\ }\href {arXiv:1802.06994v2} {\ }\Eprint
  {http://arxiv.org/abs/arXiv:1802.06994v2} {arXiv:1802.06994v2} \BibitemShut
  {NoStop}%
\end{thebibliography}%

\end{document}